\documentclass[onecolumn]{aa} 

\usepackage{graphicx}
\usepackage{float}

\usepackage{txfonts}
\usepackage{array}
\usepackage{multirow}
\usepackage{amsmath,amstext}
\usepackage{newtxmath} 
\usepackage{appendix}
\begin{document}

   \title{Deep learning Approach for Classifying, Detecting and Predicting Photometric Redshifts of Quasars in the Sloan Digital Sky Survey Stripe 82}


   \author{J. Pasquet-Itam
          \inst{1}\inst{2}
          \and
          J. Pasquet\inst{3}\inst{4}
          }

   \institute{LUPM UMR 5299 CNRS/UM, Universit\'e de Montpellier, CC 72, 34095 Montpellier Cedex 05, France
              \and
              CPPM, CNRS-IN2P3, Universit\'e Aix Marseille II, CC 907, 13288 Marseille cedex 9, France\\
                            \email{pasquet@cppm.in2p3.fr}
         \and
         LIRMM UMR 5506 - team ICAR, Universit\'e de Montpellier, Campus St Priest, 34090 Montpellier
         \and 
             LSIS UMR 7296, CNRS, ENSAM, Universit\'e De Toulon et Aix-Marseille, B\^atiment Polytech, 13397 Marseille\\
             \email{jerome.pasquet@lsis.org}
             }

             \authorrunning{J. Pasquet-Itam and J. Pasquet}
             \titlerunning{Classification of quasars based on a deep learning approach}
   \date{Accepted November 3, 2017}


  \abstract
   {We apply a convolutional neural network (CNN) to classify and detect quasars in the Sloan Digital Sky Survey Stripe 82 and also to predict the photometric redshifts of quasars. The network takes the variability of objects into account by converting light curves into images. The width of the images, noted $w$, corresponds to the five magnitudes ugriz and the height of the images, noted $h$, represents the date of the observation. The CNN provides good results since its precision is  $0.988$ for a recall of $0.90$, compared to a precision of $0.985$ for the same recall with a random forest classifier. Moreover 175 new quasar candidates are found with the CNN considering a fixed recall of $0.97$.
   The combination of probabilities given by the CNN and the random forest makes good performance even better with a precision of $0.99$ for a recall of $0.90$.
   
   For the redshift predictions, the CNN presents excellent results which are higher than those obtained with a feature extraction step and different classifiers (a K-nearest-neighbors, a support vector machine, a random forest and a gaussian process classifier). Indeed, the accuracy of the CNN within $|\Delta z|<0.1$ can reach 78.09$\%$, within $|\Delta z|<0.2$ reaches 86.15$\%$, within $|\Delta z|<0.3$  reaches 91.2$\%$ and the value of rms is 0.359. The performance of the KNN decreases for the three $|\Delta z|$ regions, since within the accuracy of $|\Delta z|<0.1$, $|\Delta z|<0.2$ and $|\Delta z|<0.3$ is 73.72$\%$, 82.46$\%$ and 90.09$\%$ respectively, and the value of rms amounts to 0.395. So the CNN successfully reduces the dispersion and the catastrophic redshifts of quasars.
   This new method is very promising for the future of big databases like the Large Synoptic Survey Telescope.}


+   \maketitle

%

\section{Introduction} \label{sec:intro}
Quasars are powered by accretion onto supermassive black holes at the dynamical centers of their host galaxies, producing high luminosities spanning a broad range of frequencies. They are of paramount importance in astronomy. For example, their studies can inform on massive blackhole (e.g. \cite{2012MNRAS.420..732P}). Moreover, as they are the most luminous Active Galactic Nuclei (AGN), they can be seen far across the Universe. So they give clues to the evolution and structure of galaxies (e.g. \cite{2006ApJS..163....1H}). They are also used as background objects to study the absorption of intergalactic matter in the line of sight, which have many applications in Cosmology (e.g. \cite{2008ApJ...679.1144L}). With the advent of large and dedicated surveys such as the Sloan Digital Sky Survey (SDSS; \cite{2000AJ....120.1579Y}) and the 2dF Quasar Redshift Survey (2QZ; \cite{2009MNRAS.392...19C}), the number of known quasars has rapidly increased. Thus, the SDSS DR7 Quasar catalog (\cite{2010AJ....139.2360S}) contains 105,783 spectroscopically confirmed quasars. The catalog covers an area of $\simeq 9380 \,\, \textrm{deg}^{2}$ and the quasar redshifts range from 0.065 to 5.46. 

With the soon coming of the Large Synoptic Survey Telescope (\cite{2009arXiv0912.0201L}), it is important to develop classification tools for quasar detection given the huge amount of future data. In this way, machine learning algorithms are being used increasingly. These algorithms permit to predict the label of an object thanks to the extraction of different features which characterize the object (e.g. the color of the source). 
Several classifiers are now commonly used in astronomy like random forests which are a set of decision trees (\cite{Quinlan86inductionof}), Naives Bayes (\cite{opac-b1102308}), Neural Networks (\cite{Rumelhart:1986:LIR:104279.104293}) and Support Vector Machines (\cite{Cortes:1995:SN:218919.218929}). These methods are very powerful in classification and detection of variable objects in astronomy (e.g. \cite{Eyer2005MNRAS.358...30E}; \cite{Dubath2011MNRAS.414.2602D}; \cite{Blomme2011MNRAS.418...96B}; \cite{Rimoldini2012MNRAS.427.2917R}; \cite{Peng2012MNRAS.425.2599P}; \cite{2015ApJ...811...95P}). We can also cite the recent work of \citet{2016ApJ...817...73H} on the classification and the detection of QSOs in the Pan-STARR S1 (PS1) $3\pi$ survey. This is a multi-epoch survey that covered three quarters of the sky at typically 35 epochs between 2010 and the beginning of 2014 with five filters ($g_{P1}$, $r_{P1}$, $i_{P1}$, $z_{P1}$, $y_{P1}$). They use a random forest classifier and colors and a structure function as features, to identify 1,000,000 QSO candidates.

The main motivation for this work is to propose a new classification and a detection method for quasars in the Sloan Digital Sky Survey Stripe 82, that can be easily adapted to large future surveys like LSST or DES (\cite{2005astro.ph.10346T}).
The algorithms mentioned above, involves a feature extraction step but the set of features can be incomplete to characterize the variability of quasars. That is why we proposed to use another branch of machine learning namely deep learning. It is a supervised learning which takes raw data into account and extracts by itself the best features for a given problem. This method gives very good results in many fields. In particular we use a Convolutional Neural Network (CNN) architecture which gives excellent results in several signal processing challenges as Imagenet (\cite{ImagenetChallenge}), LifeClef (\cite{LifeclefChallenge}) etc... This approach is very recent in Astronomy and its first applications show good results, for example for the classification of galaxy morphologies from astronomical images (\cite{2015ApJS..221....8H}).

In this work, we propose an innovative architecture based on a CNN to detect and classify quasars from light curves, thus taking into account the variability of objects. We also apply this kind of architecture to estimate the photometric redshifts of quasars. The estimation of photometric redshifts by a CNN classifier is an original method which is very promising.
This paper is organized as follows. In Section 2, we introduce the Stripe 82 data set. In Section 3, we describe the CNN architecture and processing.  In Section 4, we propose our CNN architecture for the detection and the classification of quasars. In Section 5, we analyze and discuss the new quasar candidates detected by our method. Then, we compare our algorithm with a random forest classifier and combine them. In Section 6 we propose to use a similar CNN architecture to predict the photometric redshifts of quasars. Finally we summarize our results in Section 7. 

\section{Data} \label{sec:style}

\begin{figure*}
\centering
\includegraphics[width=0.92\linewidth, trim={1.5cm 0.75cm 1.5cm 0.75cm}]{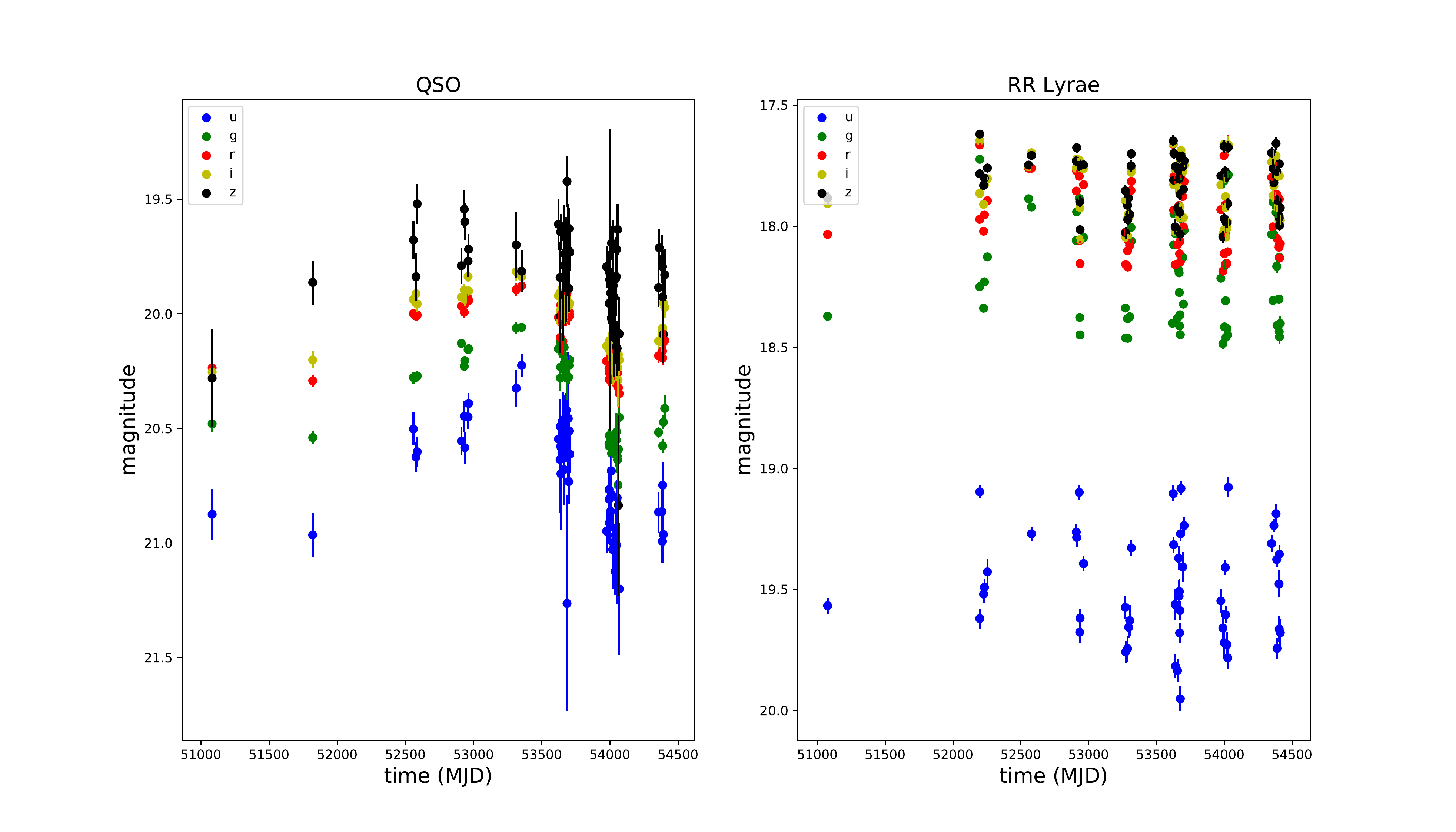}
\caption{{{Two examples of variable object types in the UWVSC catalog. On left panel, it is a quasar light curve and on right panel a RR Lyrae light curve.}}}
\label{fig:figure_lc}
\end{figure*}

\begin{figure}
\centering
\includegraphics[width=0.5\linewidth, trim={2.5cm 0.75cm 2.5cm 0.75cm}]{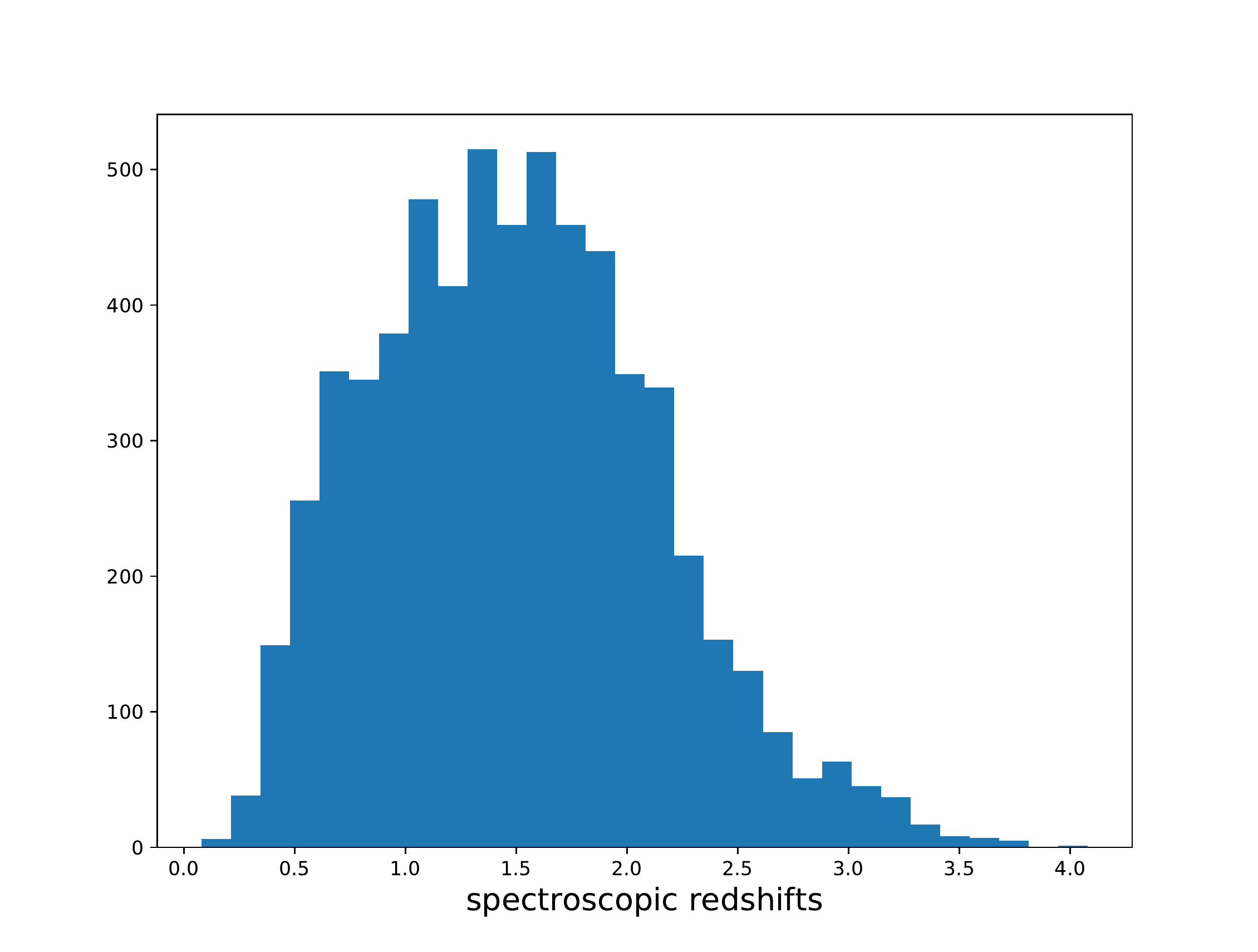}
\caption{{{Distribution of spectroscopic redshifts of known quasars in the UWVSC catalog. }}}
\label{reddis}
\end{figure}

The Sloan Digital Sky Survey (SDSS) is a multi-filter imaging and spectroscopic redshift survey using a dedicated 2.5-meter telescope at Apache Point observatory in New Mexico. It provides deep photometry ($r<22.5$) in five passbands (ugriz). The SDSS has imaged a 2.5 degree wide stripe along the Celestial Equator in the Southern Galactic Cap several times, called Stripe 82. It is a deeper survey of 275 $\textrm{deg}^{2}$. It was previously imaged about once to three times a year from 2000 to 2005 (SDSS-I), then with an increased cadence of 10-20 times a year from 2005 to 2008 (SDSS-II) as part of the SDSS-II supernovae survey (\cite{frieman2008AJ....135..338F}). There are on average 53 epochs, over a time span of 5 to 10 years (\cite{Abazajian2009ApJS..182..543A}).

The imaging data used in our work consists of objects solely from the publicly available variable source catalog (UWVSC; \cite{ivezic2007AJ....134..973I}, \cite{sesar2007AJ....134.2236S}) constructed by researchers at the University of Washington. This catalog contains 67,507 unresolved, variable candidates with $g \leq 20.5 \,\, \textrm{mag}$, at least 10 observations in both g and r bands, and a light curve with a root-mean-scatter (rms) $> 0.05$ mag and $\chi^2$ per degree of freedom $>3$ in both g and r bands. 
{{Among the data, some variable objects have been identified, they are essentially quasars and pulsating stars (see Figure \ref{fig:figure_lc}). However a large part of the data consists of unknown variable objects. In this way this catalog is interesting to identify new variable objects. }}
This catalog and all light curves are publicly available.\footnote{\url{http://www.astro.washington.edu/users/ivezic/sdss/catalogs/S82variables.html}}
We use the UWVSC as the basis for learning and testing for several reasons: 1) it contains over 9000 known spectroscopically confirmed quasars (\cite{meusinger2011A&A...525A..37M}) {{ whose the distribution of redshifts is shown in Figure \ref{reddis} }}; 2) it is a robust variable catalog with a good photometry; 3) {{the catalog is a useful testbed for time domain science to prepare future data sets like LSST. }}

\section{The Convolutional Neural Network} \label{sec:floats}
In this work, we are particularly interested in CNN which are an approach to deep learning methods that are proving their worth in many research fields. 

\subsection{Light Curve Images}
\label{LCI}

As a CNN takes images as input, we had to find a way to convert a light curve to an image taking into account the variability of objects which gives a crucial information for the classification of variable objects. Thus, we propose to create images whose the width is represented by the five magnitudes (u, g, r, i and z), and the height corresponds to the date of the observation.
In Stripe 82, there are a maximum of 3340 days of observation so images should have a dimension of 5$\times$3340 pixels. However processing these images is very costly in VRAM memory, so we divided the time interval of a light curve by averaging the observations taken on two consecutive days, so as to get images of dimensions of 5$\times$1670 pixels. Then 60 pixels were appended to the edges of the image to avoid side-effects. 
Therefore the size of the final images, called hereafter, LCI (Light Curve Images), is 5$\times$1700 pixels.

In order to increase the robustness of the network, the learning needs to be free of the positions of points. To do this, we generate new light curves by making time translations for all points on a given light curve. Thus only the global shape of the light curve is taken into account and no positions of points are considered as more important than others. This process is similar to that of classical data augmentation (\cite{dataAugmentationTimeSerie, ImageNet}) in the CNN learning and increases the size of the database by a factor of 13.

\subsection{Introduction to the CNN}

\begin{figure*}
\centering
\includegraphics[height=7.5cm]{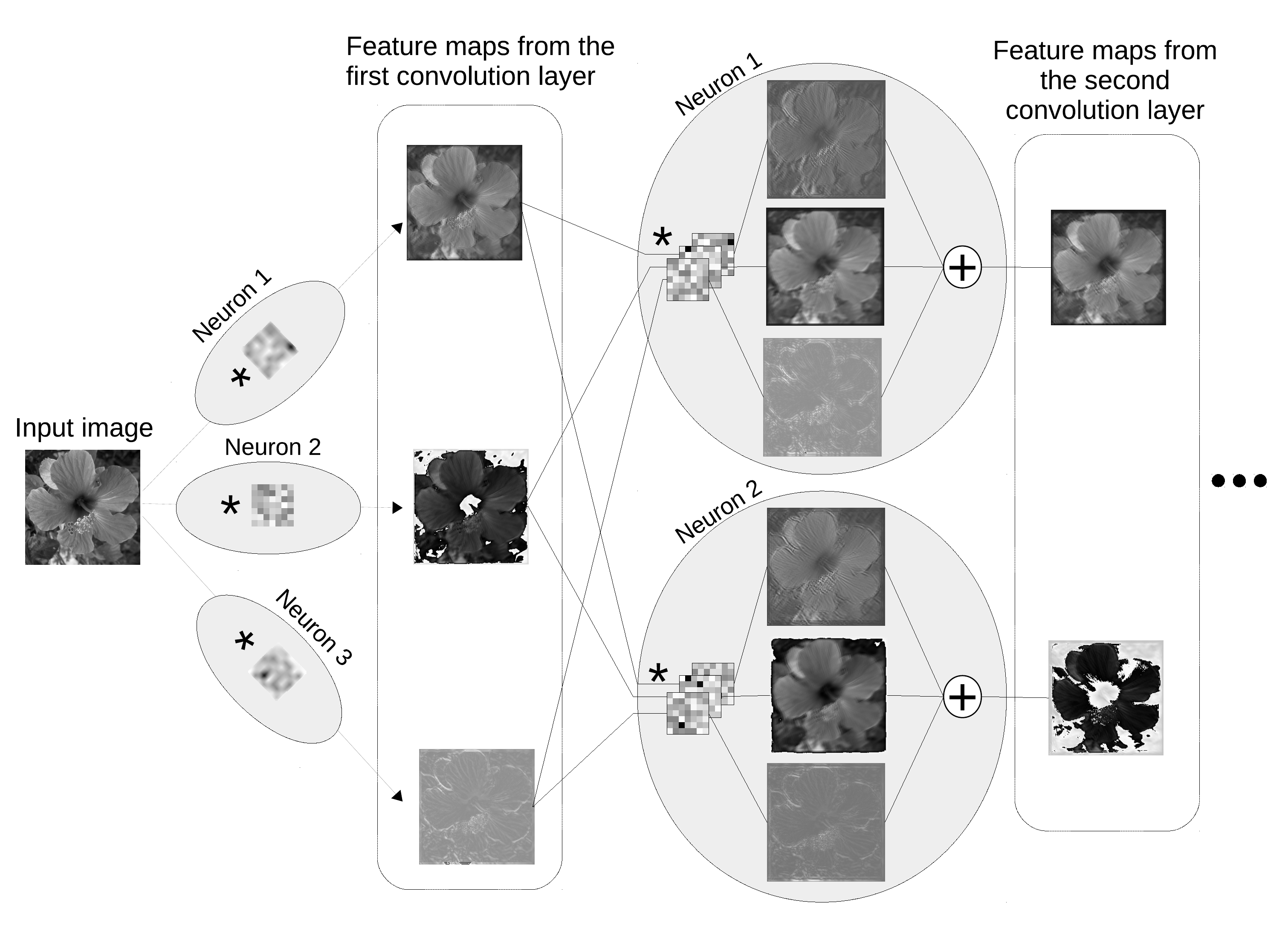}
\caption{Representation of two convolution layers of a network. The first layer is composed of 3 neurons making a convolution between the input image and their kernels. The second layer includes two neurons making a sum of convolutions as defined in the equation \ref{convolutionEq}. }
\label{convolution_schema}
\end{figure*}

An artificial neuron is a computational model inspired by natural neurons. A natural neuron is an electrically excitable cell that processes and transmits information via synapses which are connected with other cells. When the signal received is strong enough (higher than a specific threshold), the neuron is activated and emits a signal which might activate other neurons. Artificial neurons do not reproduce the complexity of real neurons, but the global structure is quite similar. Indeed, input data are multiplied by weights and then computed by a mathematical function which determines the activation of the neuron. An artificial neural network is then composed of different neuron layers connected with each other. A layer of rank $n$ takes as input the output of the layer of rank $n-1$. The nomenclature of layers is the following: i) a layer whose input is not connected to the output of another layer, but to the raw data is called input layer; ii) a layer whose output is not connected is called an output layer; iii) a layer is called "hidden" if it has either input or output.
Each layer is composed of several tens of thousands of neurons. In the specific case of a CNN, neurons perform convolution (see Section \ref{conv}) and pooling operations (see Section \ref{pool}). Layers are sequentially processed as follows: first, a convolution operation is applied on raw input data; then the output signal is modified by a non linear function; finally a pooling operation can be processed. Note that the output of a layer could be considered as a set of images. In CNN terminology, each image has named feature map.
After all convolution and pooling layers, the last convolution layer is connected to a succession of layers called "fully connected layers" and operating as a classical neural network. The last one uses a softmax operation to give a probability that the input light curve is either a quasar light curve or another object.
To perform the learning phase, parameters of convolution and fully connected layers are tuned using a stochastic gradient descent specific to the given problem, in this case the recognition of quasar light curves. This optimization process is very costly, but it can be highly parallelizable. We use the Caffe (\cite{caffe}) framework to train our CNN. The results are obtained using a GTX Titan X card, packed in 3,072 cores with a 1 GHz base.

\subsubsection{Convolution}
\label{conv}

If we consider a layer or a set of feature maps as input, the first step is to apply convolutions. For the first layer, the convolution is done between the input image and a filter. Each filter leads to a filtered image. 
Convolution layers are composed of convolutif neurons. Each convolutif neuron applies the sum of 2D convolutions between the input feature maps and its kernel. In the simple case where only one feature map is passed into the input convolutif neuron, the 2D convolution between the $\textrm{K}$ kernel of size $w\times h$ and the input feature map  $I\in \mathbb{R}^{2}$ is noted $\mathbf{I}*\mathbf{K}$ and is defined as:
$(I*K)_{x,y}=\sum_{x'=x-\frac{w}{2}}^{x+\frac{w}{2}}\,\,\sum_{y'=y-\frac{h}{2}}^{y+\frac{h}{2}}\mathbf{K}_{x'+\frac{w}{2}-x,\, y'+\frac{h}{2}-y}\mathbf{I}_{x',y'}$
with (x,y) the coordinates of a given pixel into the output feature map. 

In the case of convolutional neural networks, a neuron takes as input each of $p$ feature maps of the previously layer noted $I^{l}$ with $l\in\{0...p\}$. The resulting feature map is the sum of $p$ 2D convolutions between the kernel $K^{l}$ and the map $I^{l}$ (see Figure \ref{convolution_schema}) and is defined as:
\begin{equation}
(\mathbf{I}*\mathbf{K})=\sum_{l=0}^{p}(\mathbf{K}^{l}\mathbf{*I}^{l})
\label{convolutionEq}
\end{equation}

In this work we propose to use two types of convolutions that we call “temporal convolutions” and “filter convolutions”. The temporal convolutions use a kernel with a x-dimension of 1 pixel, so the five magnitudes (u, g, r, i and z) are convoluted separately, and with a y-dimension variable in the interval \{5, 11, 21, 41\} pixels. Thus the temporal convolutions take into account a value of magnitude at different times and at different resolutions. The advantage of this type of convolutions is to create a network which is able to detect short and long variability patterns.

The filter convolutions use a kernel with a dimension of $5 \times 1$ pixels, so they merge the values of the five magnitudes in order to integrate the information from color which is an important feature to characterize variable objects, at a given time.
\subsubsection{Pooling}
\label{pool}
The network can be composed of pooling layers which quantify the information while reducing the data volume. The Pooling Layer operates independently on each feature map. On each feature map, it slides a specific filter which represents the local distribution.  The two most used methods consist in selecting only the maximal or the mean value of the data in the local region.
As the observational data are not continuous in time, several pixels in the LCI are equal to zero. Thus we decide to adapt the pooling by the mean, i.e. we don't taking into account the null pixels in the computation of the mean. Our architecture includes this improvement of the pooling on the first pooling layers of the network. The others layers contain a max pooling.

\subsubsection{Activation Functions}
\label{activation}
The convolution layers are followed by non linear transformations whose goal is to solve the non-linear classification problems. The two most used functions are the ReLU (Rectify Lineair Unit, \cite{ReLU}) defined by $f(x)=max(x,0)$ and the hyperbolic tangent. 
In our network, to saturate the input signal we apply a hyperbolic tangent function on all of the first convolution layers. The other layers use a PReLU (\cite{PRELU}) function defined as: $f(x)=\begin{cases}
\alpha x & x<0\\
x & x\geq0
\end{cases}$
, with $\alpha$ an hyperparameter defined by back-propagation.

\section{Our CNN architecture}

\begin{figure}[h!]
\includegraphics[height=8.2cm,angle=90]{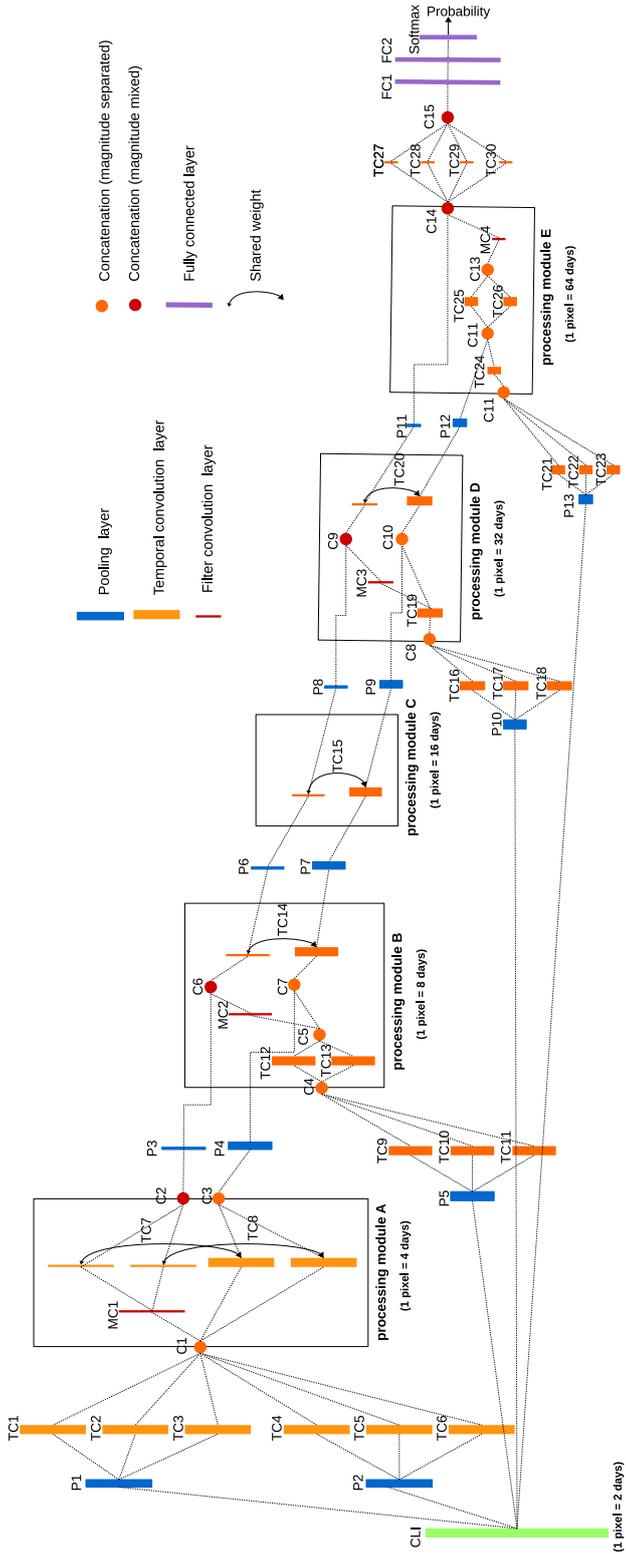}
\caption{Representation of the architecture that we are proposing. The structure is subdivided into five successive processing blocks at different temporal resolutions. Two types of convolutions are used: temporal convolutions with four kernel sizes: {$41\times1$, $21\times1$, $11\times1$ and $5\times1$} and filter convolutions with a kernel size of $5\times1$.}
\label{notre_architecture}
\end{figure}

The overall structure can be subdivided into successive processing blocks at different temporal resolutions (4, 8, 16, 32 and 64 days) as shown in Figure \ref{notre_architecture}, on five levels of depth. The initial resolution of LCI is two days per pixel. At each processing level, this resolution is reduced by a factor of 2, by a max-pooling. Moreover, each processing block is powered by a set of feature maps coming from an average-pooling retrieved a parallel on the first light curve image. These feature maps are then convoluted by three types of temporal convolutions, processing  images by three different filter sizes. This set of temporal convolutions is similar to a multi-resolutions process which is used in modern architectures like the GoogleNet network (\cite{googlenet}). The resulting feature maps are then transmitted to the processing block of the associated resolution. We note $\mathbb{F}_i$ the set of resulting feature maps transmitted to the processing block $i$. Thus as shown in Figure \ref{notre_architecture}, feature maps whose a pixel is represented by 4 days are transmitted to the module A, and those whose a pixel is represented by 8 days are transmitted to the module B, etc...

In a first step, a filter convolution (MC1 in schema \ref{notre_architecture}) is applied on $\mathbb{F}_A$. The feature maps resulting from MC1 are noted $\mathbb{F}'_A$.
We then apply two temporal convolutions (TC7 and TC8) with weights called "shared". This term can be explained by the difference of the convolution kernels applied to the sets $\mathbb{F}_A$ and $\mathbb{F}'_A$. We modify the standard convolutions so a given kernel can be applied on two sets of feature maps with different sizes. The goal is to highlight similar temporal patterns between two sets of feature maps $\mathbb{F}_A$ and $\mathbb{F}'_A$, and so between mixed and not-mixed magnitudes amongst themselves. The feature maps coming from the convolution layers TC7 and TC8 on the set $\mathbb{F}_A$ are concatenated and then pooled in the pooling layer P4. The resulting feature maps are noted $\Omega_B$ and then transmitted to the processing block B. The same process is applied to the set of feature maps $\mathbb{F}'_A$ giving the set $\Omega'_B$.

The second processing block performs a temporal convolution with two kernels of different sizes (TC12 and TC13) on $\mathbb{F}_B$. The resulting feature maps are then transmitted to two layers: C7 and MC2. In the C7 layer, they are concatenated with the set $\Omega_B$. In the MC2 layer, they are convoluted by a filter. The resulting feature maps are concatenated with that of the set $\Omega'_B$. The set of produced feature maps are temporally convoluted with shared weights in the TC14 layer. The result is then pooled and transmitted to the processing block C.

The functioning of the rest of the blocks are similar to that of block B. The number of feature maps and the size of each layer are noted in Table \ref{table_reseau} in the appendix. We can remark that the processing block E transmits only feature maps whose magnitudes are merged, which are then temporally convoluted and transmitted to the fully connected layers. After convolution layers, the size of feature maps are 53$\times$1 pixel. 

In the network architecture we use two processes to avoid over fitting. First, all the feature maps are normalized using the batch normalization (\cite{batchnorm}). Second, the outputs of the fully connected layers are randomly dropout (\cite{dropout}). During the back-propagation processing, the network has to determine a large number of parameters, namely 1,802,032 in the convolution layers and 11,468,80 in the fully-connected layers.

\section{Classification Results}

\subsection{Experimental protocol}

We did five cross-validations of the database by always selecting 75$\%$ of the LCI for the learning base and 25$\%$ for the testing base.For each of the five cross-validations, each CNN completed its learning on 60 epochs (during an epoch, each LCI is transmitted to the network and its error is back-propagated). {{Each CNN has three outputs on the softmax layer corresponding to the following classes : quasars, pulsating stars (RR Lyrae and $\delta$ Scuti) and other objects}}. During the testing phase, each CNN gives a list of detected quasars in the testing base. We merge the lists  given by each CNN into one list that we evaluate.

\subsection{Results}
\begin{figure*}[h!]
\includegraphics[width=\linewidth, trim={4cm 1cm 4cm 1cm}]{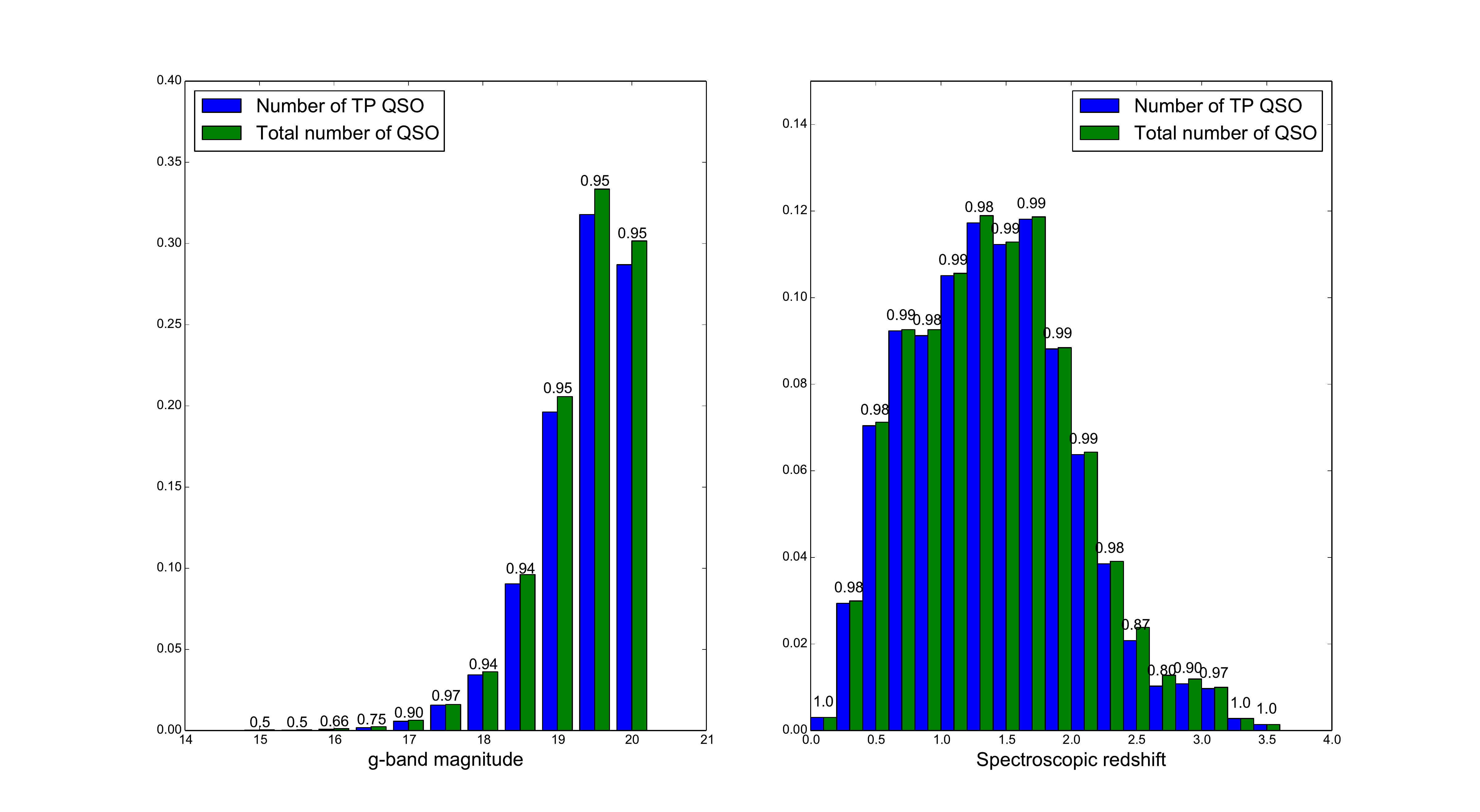}
\caption{{{The left histogram illustrates the performance of the CNN depending on the median g-band magnitude. The right histogram shows the performance of the CNN in function of the redshift. On each histogram the blue bars represent the number of well detected quasars by the CNN and the green bars the total number of quasars inside the corresponding bin. The recall is indicated over each couple of bars. }}}
\label{ref_1}
\end{figure*}
{{
The performance of the CNN is given on Figure \ref{ref_1} in function of the magnitude and the redshift. We can notice that for a g-band magnitude below 17 magnitudes, the value of the recall decreases until a recall of 50\%. It is due to the too low number of examples of very bright quasars in the training set. Indeed there are only 22 light curves of quasars in the training database with magnitudes between 15 and 17. However the recall is similar whatever magnitudes above 17 for the g-band magnitude. It is a very interesting result because it means that the CNN performance does not depend on the magnitude but only on the number of objects in the training database. 
This effect is less visible on the right histogram of Figure \ref{ref_1}. Indeed, it is enough to consider only 5\% of the training database to reach a recall between 98\% and 99\%. This experiment shows that the CNN is invariant to redshift. 
}}

\begin{figure}[h!]
\includegraphics[height=7.2cm]{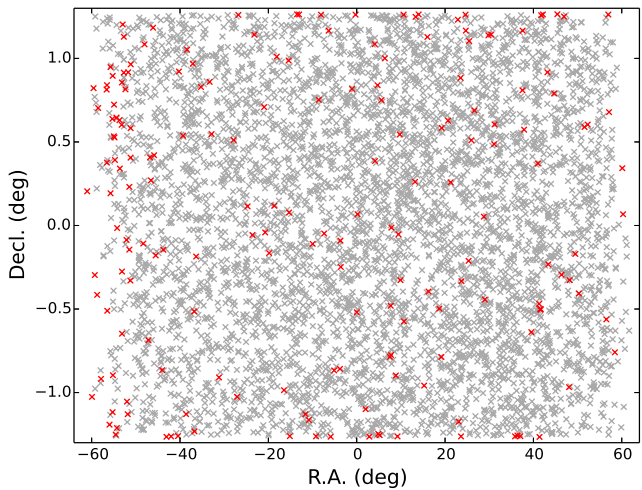}
\caption{Spatial distribution of quasars detected by the CNN during the testing phase. The gray crosses represent the  well-known detected quasars, and the red crosses are the 175 new quasar candidates. They are uniformly distributed in the sky.}
\label{spatial}
\end{figure}

In the testing base, for a fixed recall of 0.97, 175 new quasars detected by the CNN have never been identified before. We call them quasar candidates. Figure \ref{spatial} represents the spatial distribution of found quasars in the testing base by the CNN. The red crosses characterize the new quasar candidates.

\begin{figure}[h!]
\includegraphics[height=7.2cm]{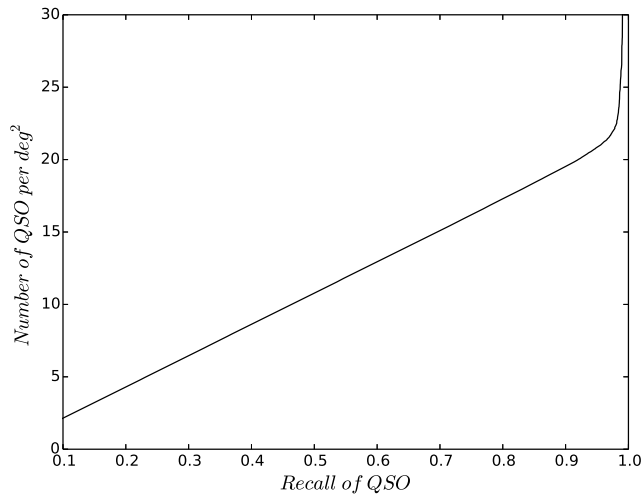}
\caption{Average number of quasars in the sky per square degree, detected by the CNN, against the recall. The larger the recall, the higher the number of the detected quasars is.  For a recall of 0.92, the average number of detected quasars is 20 per square degree. Then this number is drastically increased since the precision is reduced and so the contamination of non-quasar sources is increased.}
\label{qso_par_degree}
\end{figure}

As we can see, the quasars detected by the CNN are distributed in an uniform manner. Figure \ref{qso_par_degree} shows the average number of quasars in the sky per square degree, detected by the CNN, against the recall.  As the recall increases, the number of quasars per square degree increases, which is consistent as we detected more and more quasars. For a recall around 0.92, the average number of quasars per square degree is about 20. Then, this number is drastically increased because the precision is reduced and the sample is contaminated by sources which are not quasars.

\begin{figure*}
\centering
\includegraphics[width=18.2cm]{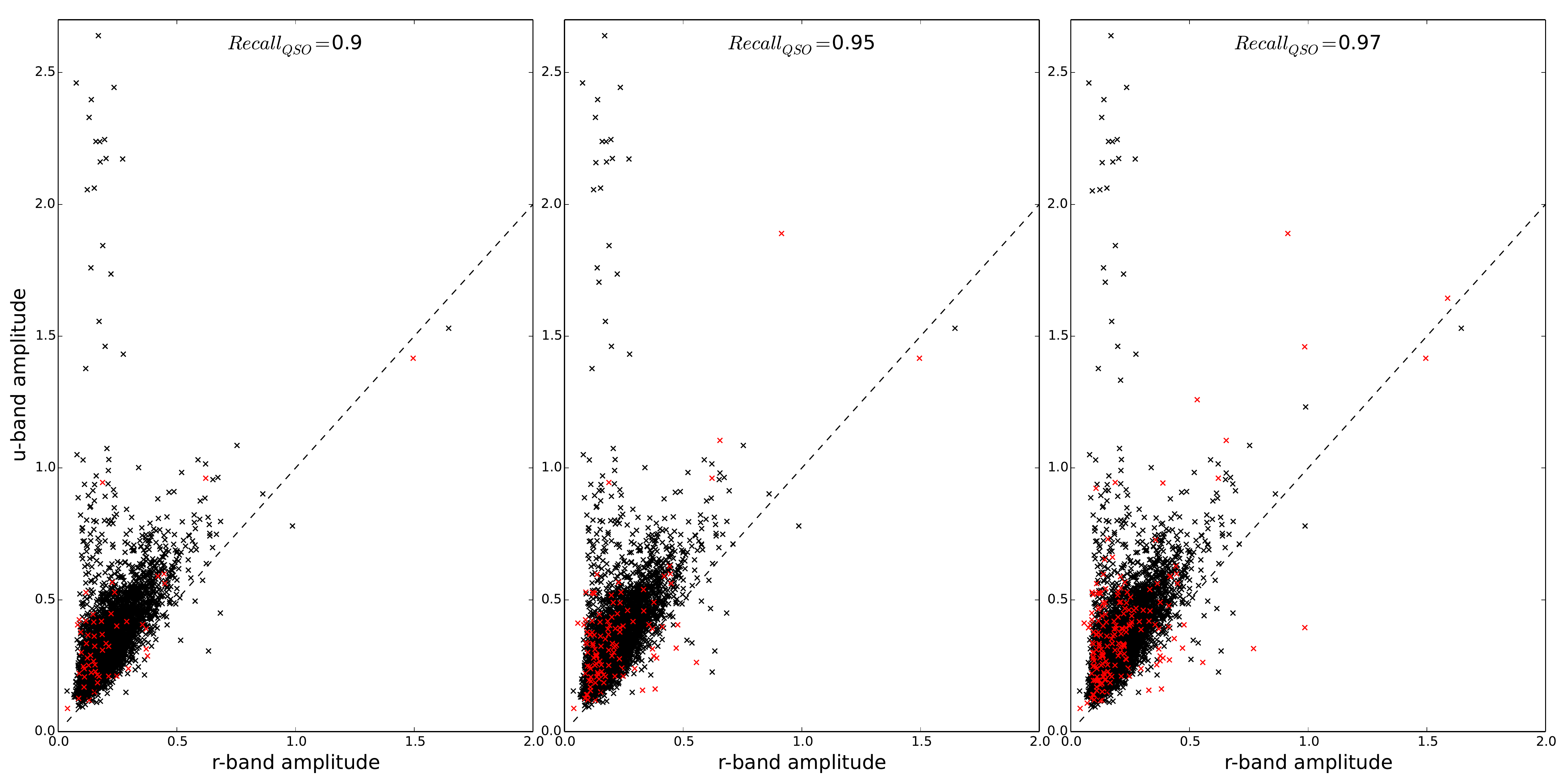}
\caption{Amplitudes of variation of quasars detected by the CNN from the u-band filter against those from the r-band filter at three different recalls: 0.90, 0.95 and 0.97. The black crosses represent all the known quasars during the testing phase. The red crosses are the 175 new quasar candidates. The dashed line represent the line $y=x$. The quasars show larger variation amplitudes in bluer-bands. This tendency highlights a strong wavelength dependence.
}
\label{ampli}
\end{figure*}

It is also interesting to highlight that a well known property of quasars is met by the new quasar candidates, namely a "bluer when brighter" tendency. This trend has been well established in the UV/optical color variations in quasar (e.g. \cite{cristiani1997A&A...321..123C}, \cite{giveon1999MNRAS.306..637G}, \cite{vanden2004ApJ...601..692V}).  Figure \ref{ampli} represents the amplitude of variations of detected quasars in the u-band filter against the r-band filter at different recalls. We note that $83.6\%$ and $88.7\%$ of variation amplitudes in the u-band filter are larger than in the r-band filter for a recall of 0.90 and 0.97 respectively. Thus the detected quasars show larger variation amplitudes in bluer bands and so a strong wavelength dependence.

\subsection{Comparison with a random forest classifier}
We compare the performance of our algorithm with that of a random forest classifier whose we empirically estimated the best parameters on the same database. It contains 400 decision trees and an unlimited depth. The features used are included in a python library named FATS (Feature Analysis for Time Series, \cite{Nun2015arXiv150600010N}) which is a compilation of some of the existing light-curve features.

At a fixed recall of 0.90, the precision is 0.988 for the CNN and 0.985 for the random forest. Then for a fixed recall of 0.97, the precision is 0.964 and 0.973 for the CNN and the random forest respectively. The performances of the two methods are closed and a little better for the random forest. A possible explication concerns the number of freedom degrees. Indeed for the random forest, about 640 000 parameters are defined whereas for the CNN there are about 13 000 000. Thus, due to the large number of parameters that have to be determined by backpropagation, the CNN should have better performance with more data, especially with large surveys.

\subsection{Combination of a CNN and a random forest}
\begin{figure} 
\includegraphics[height=7.4cm]{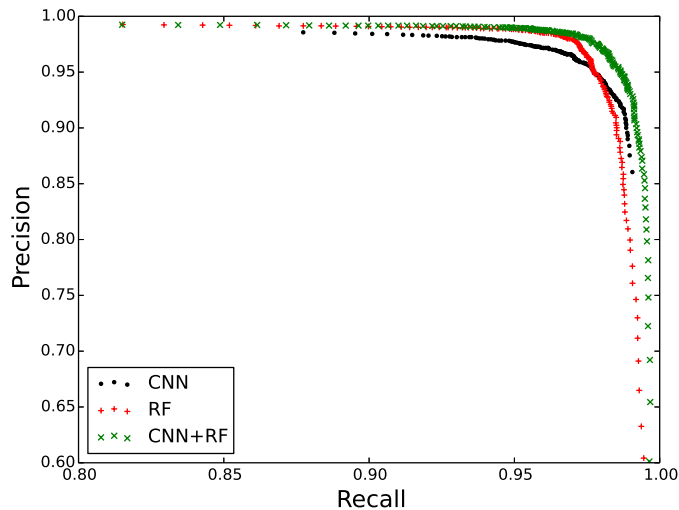}
\caption{ROC curve which plots the precision of the classifier against the recall. The performance of the CNN classifier is represented by black dots, those of the random forest by red plus and the performance of the combination of the two classifiers is represented by green crosses. }
\label{ROC}
\end{figure}

We combine the probabilities given by the CNN and the random forest by averaging them. For a fixed recall of 0.90 the precision is  0.99 and for a recall of 0.97, the precision is 0.98. Thus the combination of the two classifiers makes good performance even better. Figure \ref{ROC} shows the receiver operating characteristic curve (ROC curve hereafter) which is a graphical plot that illustrates the performance of a classifier by plotting the precision against the recall. We can see that the random forest performance (red curve) is better than that of the CNN (black curve) until a recall of 0.978, where the CNN performance slightly drops. Moreover the ROC curve representing the combination of the two classifiers (green curve) is above the two others and shows that combining a CNN classifier and a random forest classifier gives better classification performance.
{{
The improvement obtained by the combination of the CNN and the random forest can be explained by the complementarity between the features given to the random forest and the features extracted by the CNN. 
Indeed, features used by the random forest are defined by the user and are specific for the classification of light curves of variable objects in general but they could not be perfectly designed for this classification problem that we considered. 
On another side the CNN learns from scratch without any prior. The CNN found relevant features which are specific to the used database and so complete the information given by the features used by the RF. However, since the CNN learns features from the data, if there is not a large number of examples for a kind of objects, such as the high redshift quasars, the CNN does not find and learn the best features. In this case, it is relevant to use the results from the random forest to improve the classification. So depending on the number of data, the random forest features or the CNN features can complete each other.
}}

\section{Photometric redshifts of quasars}

Photometric redshifts are a way to determine the redshift of an object by using only the apparent magnitudes through different filters, or photometric images. They constitute a powerful technique because they allow us to be free of spectroscopy data which are limited by the brightness of the source and by the cost of instruments. This methodology has been developed by Baum (\cite{1962IAUS...15..390B}) by observing the Spectral Energy Distribution (SED) of six elliptic galaxies in the Virgo cluster in nine bands from 3730$\text{\AA}$ to 9875$\text{\AA}$. The approach using template-fitting models which extracts features from celestial observational information and then matches them with the designed templates constructed by theoretical models or real observations has been used intensively (e.g. \cite{bolzonella2000A&A...363..476B}, \cite{ilbert2009A&A...500..981C} \cite{ilbert2010ApJ...709..644I}). However the accuracy of the method strongly depends on simulated or real data. Moreover, the emergence of massive photometric data obtained by multiple large-scale sky surveys suggests the need of an automatic method such as machine learning algorithms. Several methods were used to estimate photometric redshifts of galaxies or quasars like a K-nearest neighbors (e.g. \cite{zhang2013AJ....146...22Z}, \cite{kugler2015A&A...576A.132K}), an artificial neural network (e.g. \cite{firth2003MNRAS.339.1195F}, \cite{collister2004PASP..116..345C}, \cite{blake2007MNRAS.374.1527B}, \cite{oyaizu2008ApJ...674..768O}, \cite{yeche2010A&A...523A..14Y}, \cite{zhang2009MNRAS.392..233Z}), both a K-nearest neighbors and a support vector machine (\cite{han1674-4527-16-5-005}).

We propose to predict the photometric redshifts of quasars with a CNN. {{For that, we use 80\% of the quasar light curves for the training database and 20\% for the testing database. To reduce the variability we cross validate the experiment and only show the mean of the results.}} The distribution of known spectroscopy redshifts are sliced in 60 bins of 0.04 in width. We used a network with a similar architecture that is represented in Figure \ref{notre_architecture}. The softmax gives the probability of belonging to each redshifts class. To predict the final regression value, results of each class are added by weighting them by the probability given by the Softmax. Again, the network takes the LCI as input (see Section \ref{LCI}) so as to include the information of the variability of objects in the estimation of redshifts.

 To evaluate the proposed method, we compare it with a more classical approach using an extraction of features. For that, we compared the performances of four classifiers namely a K-nearest neighbors (KNN), a support vector machine (SVM, with linear and Gaussian kernels), a random forest (RF) and a Gaussian process classifier.
 \begin{table}[H]
 \begin{tabular}{|>{\centering}m{3.5cm}|c|c|c|}
 \hline 
 Feature & Absolute error & $\chi ^2$ & best K\tabularnewline
 \hline 
 \hline 
 Mean & 0.282 & 0.239 & 2\tabularnewline
 \hline 
 Mean+error & 0.283 & 0.240 & 2\tabularnewline
 \hline 
 Mean+color+error & 0.263 & 0.199 & 3\tabularnewline
 \hline 
 \centering\multirow{2}{3.5cm}{\centering Mean+color amplitudes+error} & \multirow{2}{*}{0.253} & \multirow{2}{*}{0.182} & \multirow{2}{*}{4}\tabularnewline
  &  &  & \tabularnewline
 \hline 
 color+error & 0.226 & 0.163 & 4\tabularnewline
 \hline 
 color & 0.226 & 0.156 & 6\tabularnewline
 \hline 
 \end{tabular}

 \caption{Evaluation of the K-Nearest Neighbors classifier efficiency using different features based on the absolute error and the error given by a  $\chi ^2$ test. K is the number of neighbors taking into account by the KNN algorithm.}
 \label{feature_knn}
 \end{table}
 For each of these classifiers, we used the best combination of features among the mean of magnitudes, the magnitude errors, the amplitude of magnitudes, the colors and all characteristics included in the python library FATS. In the evaluation, the best results are obtained by using a KNN and only the color as a characteristic. Indeed as we can see in Table \ref{feature_knn} a learning phase with only the color as a feature shows the lower absolute error of 0.226 and the lower residual of the $\chi^2$ test with a value of 0.156. In this case, the number of neighbors taking into account, indicated by the number K in the Table \ref{feature_knn}, is equal to 6. Thus we use the performance of the KNN by extracting only the color to be compared to the performance of the CNN (see Table \ref{feature_knn}).

\begin{table*}
\centering
\begin{tabular}{|>{\centering}p{2cm}|c|c|c|c|}
\hline 
 &  {\small{$|\Delta z|<0.1$ (\%)}} & {\small{$|\Delta z|<0.2$  (\%)}} & {\small{$|\Delta z|<0.3$ (\%)}} & {\small{RMS}} \tabularnewline
\hline 
\hline 
{\small{CNN}} & {\small{79.32}} &  {\small{86.64}} & {\small{91.69}} & {\small{0.352}} \tabularnewline
\hline 
{\small{KNN}}  & {\small{73.72}} & {\small{82.46}} & {\small{90.09}} & {\small{0.395}} \tabularnewline
\hline 
\textbf{\small{KNN+CNN}} & \textbf{80.43} & \textbf{87.07} & \textbf{91.75} & \textbf{\small{0.349}} \tabularnewline
\hline 
\end{tabular}

\caption{Comparisons of the accuracy and the dispersion obtained with the CNN, the KNN and the merge of the KNN and the CNN, by computing the percentages in different $|\Delta z|$ ranges and the Root Mean Square (RMS).}

\label{table_erreur}
\end{table*}

\begin{figure*} 
\hspace{-2.3cm}
\includegraphics[width=22.5cm]{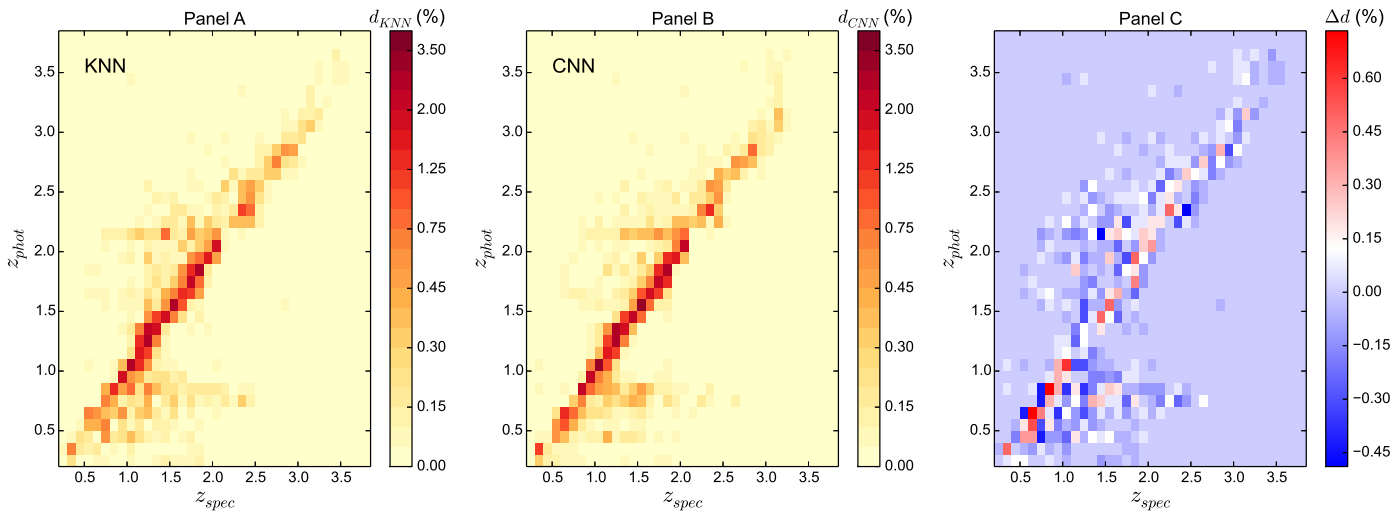} 
\caption{Panels \textbf{A} and \textbf{B} compare the photometric redshifts predicted by the KNN and the CNN  respectively against the spectroscopic redshifts. The color indicates the density of quasars in percentages. Redder the color, higher the density of quasars is. The line $y=x$ is red which means that the density of quasars is the highest and so the two methods well estimate most of the photometric redshifts compared to spectroscopic redshifts.
Panel \textbf{C} is the difference in percentages between the density of quasars given by the CNN and the density of quasars obtained by the KNN, noted $\Delta d$. In other words, when $\Delta d$ is positive (resp. negative), the color is red (resp. blue), and it means that the density of quasars given by the CNN (resp. KNN) is higher than those obtained by the KNN (resp. CNN). 
}
\label{redshift}
\end{figure*}

Figure \ref{redshift} compares photometric redshifts predicted by the KNN (panel A) and the CNN (panel B) against the spectroscopic redshifts of around 9000 quasars. The color indicates the density of quasars in percentages. We note  $d_{CNN}$ and  $d_{KNN}$  the density of quasars in percentages given by the CNN and the KNN approaches respectively. Redder the color, higher the density of quasars is. For the two methods, the density of quasars is the highest on the line y=x, showing that the most of photometric redshifts are well estimated by the two classifiers. We remark that the density is the highest for redshifts below 2.5, since the database contains a small number of high redshifts, less than 10$\%$ which is then divided between the training and the testing databases.

Panel C in Figure \ref{redshift} compares the two approaches in the estimation of the photometric redshifts since it represents the difference between the density in percentages, $d_{CNN}$ and  $d_{KNN}$ noted $\Delta d$. When the value of $\Delta d$ is positive (positive values are represented by the red color on the plot), the density of quasars given by the CNN is higher than those obtained by the KNN. Contrariwise when the value of  $\Delta d$ is negative (negative values are represented by the blue color on the plot) the density of quasars given by the KNN is higher than those given by the CNN. We can see that the line $y=x$ appears in red color, so the values of $\Delta d$ are positives showing that the density of quasars obtained by the CNN is higher than those given by the KNN and so that the CNN better predicts redshifts equals to spectroscopic redshifts than the KNN. On the contrary, the regions around the line $y=x$ are in blue meaning that the KNN has a higher error rate than the CNN and predict more catastrophic redshifts.

\begin{figure}
	\includegraphics[width=9.2cm]{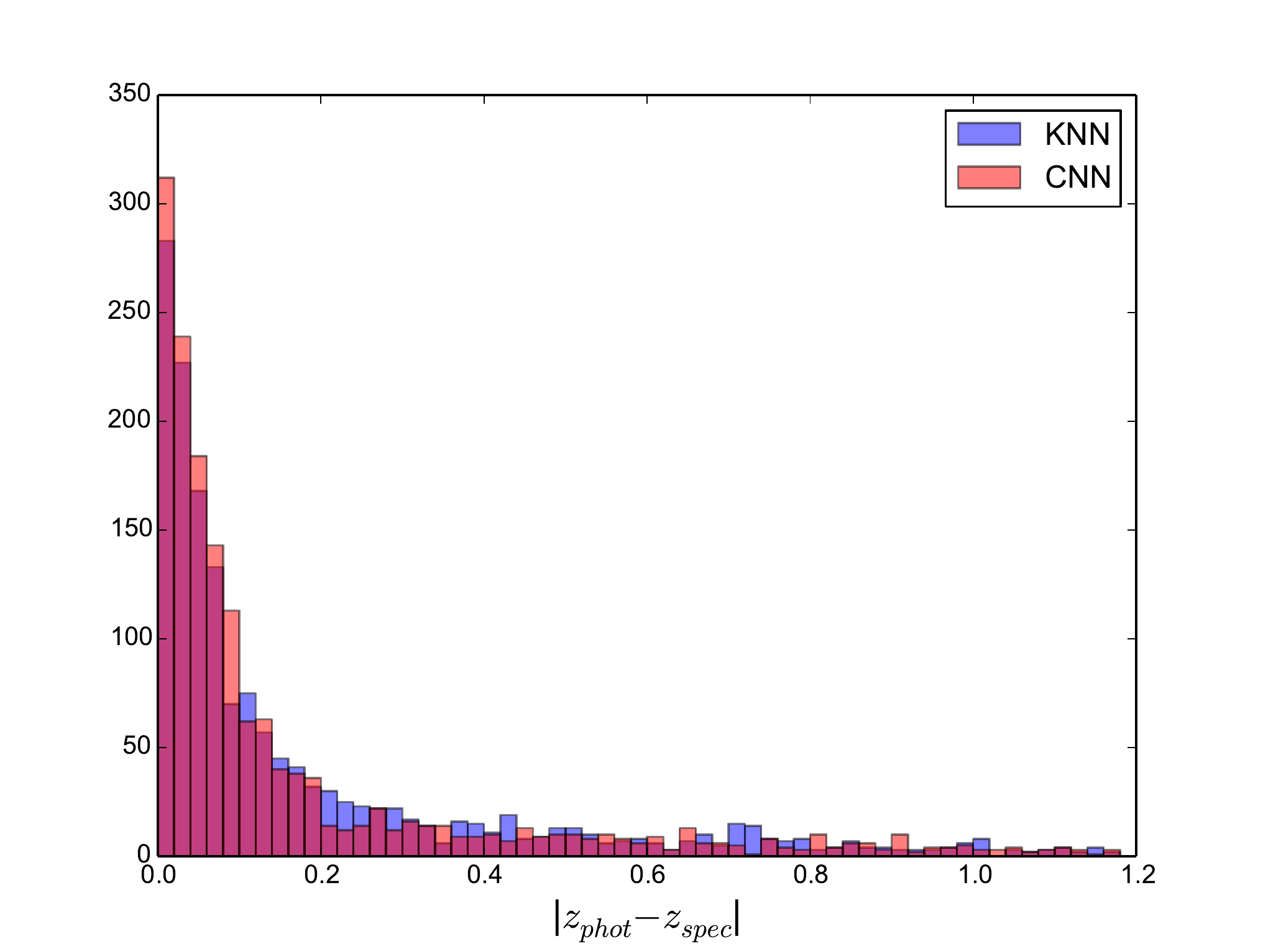}
	\caption{The blue and red histograms represent the distribution of the absolute error for the estimation of photometric redshifts by using a KNN and a CNN classifiers respectively. The number of catastrophic redshifts is reduced with the CNN as the percentage of redshift estimation with an absolute error higher than 0.1 is about 45.78\% for the KNN and 38.03\% for the CNN.  }
	\label{histo}
\end{figure}

The better accuracy of the CNN is also visible if we compare the distribution of the absolute error in the estimation of photometric redshifts of the two methods (see Figure \ref{histo}). Indeed we can see that the histogram is narrower for the estimations of photometric redshifts made by the CNN (red histogram) than those made by the KNN (blue histogram).
In addition the percentage of redshift estimations with an absolute error higher than 0.1, 0.2 and 0.3 are respectively of 38.03\%, 24.06\% and 19.07\% for the CNN; for the KNN they are of 45.78\%, 30.04\% and 22.89\%. Thus the number of catastrophic photometric redshifts is significantly reduced with the CNN. 

We also define two quantities frequently used to evaluate accuracy and dispersion of the used method that are the percentages in different $|\Delta z|$ ranges defined as:

\begin{equation}
\Delta z=\frac{z_{spec}-z_{phot}}{1+z_{spec}}
\end{equation}
and the Root Mean Square (RMS) of $|\Delta z|$ to test our redshifts prediction approach.
The better accuracy of the CNN is confirmed (see Table \ref{table_erreur}) since the proportions of $|\Delta z|$ are equals to 78.09$\%$, 86.15$\%$, 91.2$\%$ for the CNN, against 73.72$\%$, 82.46$\%$, 90.09$\%$ for the KNN. This means that more photometric redshifts are estimated with a low error by the CNN than the KNN. In addition, the dispersion of photometric redshifts is also lower with the CNN, since the RMS is 0.352 for the CNN against 0.395 for the KNN.

However the CNN has worse performance than the KNN for  the prediction of redshifts higher than 2.5 (which is visible on Panel C in Figure \ref{redshift}). It is due to the small number of redshifts higher than 2.5 in the database. There are only 600 quasars with $z_{spec}>2.5$ in the learning database and the CNN needs a lot of examples to converge.

{{To solve this problem, after the training of the KNN and the CNN we combine these two approaches in a KNN+CNN architecture. The final prediction of the KNN+CNN architecture will depend on the redshift predicted by the KNN model. If the KNN predicts a redshift higher than 2.5, this prediction is used as the final prediction. Otherwise, the prediction given by the CNN is used as the final prediction. We note $p_{KNN}$ and $p_{CNN}$  the prediction given by the KNN and the CNN respectively. The prediction given by the KNN+CNN architecture, noted $p_{KNN+CNN}$ is defined as:
\begin{equation}
p_{KNN+CNN}=\begin{cases}
p_{KNN} & \textrm{if } p_{KNN}>2.5\\
p_{CNN} & \textrm{Otherwise}\\
\end{cases}
\end{equation}
So, when the CNN does not have enough examples to learn a robust model, i.e for the high redshift estimations, the KNN model is used.
}}

The performance given by the KNN+CNN architecture is a very interesting results as shown in Table \ref{table_erreur}. Indeed, the combination of the two classifiers reduces the number of catastrophic redshifts and the dispersion, since the proportions of $|\Delta z|$ are now equal to 80.43$\%$, 87.07$\%$, 91.75$\%$ and the value of RMS is 0.349.

\section{Conclusions}
First, we have presented an original method based on a convolution neural network to classify and identify quasars in Stripe 82. The network takes the Light Curve Images as input which are built from light curves of each object in the five ugriz filters, so as to include both the crucial information of the variabiliy and the colors in the learning of the network.
The CNN classifier presents good results for the classification of quasars with a precision of 0.988 at a fixed recall of 0.90. For the same recall, the precision given by a random forest (RF) is 0.985. The very promising result is obtained by the combination of the CNN and the RF giving precisions of 0.99 for a recall of 0.90. 

Then, during the testing phase 175 new quasar candidates were detected by the CNN, with a fixed recall of 0.97. They are uniformly spatially distributed and they validate the tendency "bluer when brighter". 

Finally, we have used a CNN to predict the photometric redshifts of quasars. The performance of the CNN is higher than that of the KNN at redshifts below 2.5 with the best parameters determined experimentally.  Indeed, the proportions of $|\Delta z|$ and rms error of predicted photometry redshifts are 78.09$\%$, 86.15$\%$, 91.2$\%$ and 0.359 for the CNN; for the KNN they are 73.72$\%$, 82.46$\%$, 90.09$\%$ and 0.395. 
The number of catastrophic redshift is also reduced by using a CNN, since the number of photometric redshifts with an absolute error higher than 0.1 is about 38.03$\%$ for the CNN against 45.78$\%$ for the KNN.
Moreover the combination of a CNN and a KNN is a very promising method which better estimates redshifts higher than 2.5 and reduces the dispersion and the number of catastrophic redshifts.

Several improvements can be made for further studies. The most trivial is to use another catalog with a larger amount of data, because Deep Learning usually shows better results when there is more information. The second improvement consists of not dividing by averaging observations taken on two consecutive days, during the creation of the LCI. Indeed it is an approximation needed to reduce the computational cost, but it could be interesting to evaluate its impact on the results. Another interesting improvement is to take the errors into account in the learning phase which could show important information.

In conclusion we wish to emphasis that the development of a method able to estimate well the photometric redshifts using only photometric information is essential for the future of big databases like LSST. Understanding that Deep Learning is more and more efficient as the size of the data increases, the future of this method is very promising.

\bibliographystyle{aa}
\bibliography{biblio.bib}

\begin{appendices}
\section{Appendix}
\begin{table*}[t]
\setcounter{table}{0}
\renewcommand{\thetable}{A\arabic{table}}
\centering
\begin{tabular}{|>{\centering}p{3cm}|>{\centering}p{3cm}|c|c|c|}
\hline 
{\small{Layers}} & {\small{Inputs}} & {\small{Kernel size}} & {\small{$h\times w$}} & {\small{\#feature maps}}\tabularnewline
\hline 
\hline 
{\small{P1, P2}} & LCI & {\small{}}%
\begin{tabular}{c}
{\small{$5\times1$, $11\times1$}}\tabularnewline
{\small{(stride 2)}}\tabularnewline
\end{tabular} & {\small{$850\times5$}} & {\small{1, 1}}\tabularnewline
\hline 
{\small{TC1, TC2, TC3}} & {\small{P1}} & {\small{$11\times1$, $21\times1$, $41\times1$}} & {\small{$850\times5$}} & {\small{16, 16,16}}\tabularnewline
\hline 
{\small{TC4, TC5, TC6}} & {\small{P2}} & {\small{$11\times1$, $21\times1$, $41\times1$}} & {\small{$850\times5$}} & {\small{16, 16, 16}}\tabularnewline
\hline 
{\small{MC1}} & {\small{C1}} & {\small{$1\times5$}} & {\small{$850\times1$}} & {\small{96}}\tabularnewline
\hline 
{\small{TC7, TC8}} & {\small{MC1 and C1}} & {\small{$11\times1$, $21\times1$}} & {\small{$850\times5$ or $850\times1$}} & {\small{24, 24}}\tabularnewline
\hline 
{\small{P3}} & {\small{C2}} & {\small{$3\times1$ (stride 2)}} & {\small{$425\times1$}} & {\small{48}}\tabularnewline
\hline 
{\small{P4}} & {\small{C3}} & {\small{$3\times1$ (stride 2)}} & {\small{$425\times5$}} & {\small{48}}\tabularnewline
\hline 
{\small{P5}} & {\small{LCI}} & {\small{$21\times1$ (stride 4)}} & {\small{$425\times5$}} & {\small{1}}\tabularnewline
\hline 
{\small{TC9, TC10, TC11}} & {\small{P5}} & {\small{$5\times1$, $11\times1$, $21\times1$}} & {\small{$425\times5$}} & {\small{16,16,16}}\tabularnewline
\hline 
{\small{TC12, TC13}} & {\small{C4}} & {\small{$11\times1$, $21\times1$}} & {\small{$425\times5$}} & {\small{24, 24}}\tabularnewline
\hline 
{\small{MC2}} & {\small{C5}} & {\small{$1\times5$}} & {\small{$425\times1$}} & {\small{48}}\tabularnewline
\hline 
{\small{TC14}} & {\small{C6 and C7}} & {\small{$11\times1$}} & {\small{$425\times1$ or $425\times5$}} & {\small{96, 96}}\tabularnewline
\hline 
{\small{P6}} & {\small{TC14}} & {\small{$3\times1$ (stride 2)}} & {\small{$212\times1$}} & {\small{96}}\tabularnewline
\hline 
{\small{P7}} & {\small{TC14}} & {\small{$3\times1$ (stride 2)}} & {\small{$212\times5$}} & {\small{96}}\tabularnewline
\hline 
{\small{TC15}} & {\small{P6 and P7}} & {\small{$11\times1$}} & {\small{$425\times1$ or $425\times5$}} & {\small{48, 48}}\tabularnewline
\hline 
{\small{P8}} & {\small{TC15}} & {\small{$3\times1$ (stride 2)}} & {\small{$106\times1$}} & {\small{48}}\tabularnewline
\hline 
{\small{P9}} & {\small{TC15}} & {\small{$3\times1$ (stride 2)}} & {\small{$106\times5$}} & {\small{48}}\tabularnewline
\hline 
{\small{P10}} & {\small{LCI}} & {\small{$41\times1$ (stride 16)}} & {\small{$106\times5$}} & {\small{1}}\tabularnewline
\hline 
{\small{TC16, TC17, TC18}} & {\small{P10}} & {\small{$5\times1$, $11\times1$, $21\times1$}} & {\small{$106\times5$}} & {\small{16, 16, 16}}\tabularnewline
\hline 
{\small{TC19}} & {\small{C8}} & {\small{$11\times1$}} & {\small{$106\times5$}} & {\small{48}}\tabularnewline
\hline 
{\small{MC3}} & {\small{TC19}} & {\small{$1\times5$}} & {\small{$106\times1$}} & {\small{48}}\tabularnewline
\hline 
{\small{TC20}} & {\small{C9 and C10}} & {\small{$11\times1$}} & {\small{$106\times1$ or $106\times5$}} & {\small{128, 128}}\tabularnewline
\hline 
{\small{P11}} & {\small{TC20}} & {\small{$3\times1$ (stride 2)}} & {\small{$53\times1$ }} & {\small{48}}\tabularnewline
\hline 
{\small{P12}} & {\small{TC20}} & {\small{$3\times1$ (stride 2)}} & {\small{$53\times5$ }} & {\small{48}}\tabularnewline
\hline 
{\small{P13}} & {\small{LCI}} & {\small{$61\times1$ (stride 32)}} & {\small{$53\times5$ }} & {\small{1}}\tabularnewline
\hline 
{\small{TC21, TC22, TC23}} & {\small{P13}} & {\small{$5\times1$, $11\times1$, $21\times1$}} & {\small{$53\times5$ }} & {\small{16, 16, 16}}\tabularnewline
\hline 
{\small{TC24}} & {\small{C11}} & {\small{$11\times1$}} & {\small{$53\times5$ }} & {\small{64}}\tabularnewline
\hline 
{\small{TC25, TC26}} & {\small{C12}} & {\small{$11\times1$, $21\times1$}} & {\small{$53\times5$}} & {\small{64, 64}}\tabularnewline
\hline 
{\small{MC4}} & {\small{C13}} & {\small{$1\times5$}} & {\small{$53\times1$}} & {\small{64}}\tabularnewline
\hline 
{\small{TC27, TC28, TC29, TC30}} & {\small{C14}} & {\small{$5\times1$, $11\times1$, $21\times1$, $41\times1$}} & {\small{$53\times1$}} & {\small{48, 48, 48, 48}}\tabularnewline
\hline 
{\small{FC1, FC2}} & {\small{C15, FC1}} & {\small{-}} & {\small{- }} & {\small{1024, 1024}}\tabularnewline
\hline 
\end{tabular}
\caption{Characteristics of each layer of the CNN architecture: the name of the layer, the input layer, the size of the convolution kernel (in pixels), the size in pixels (height$\times$width) of resulting feature maps and the number of resulting feature maps. The concatenation layers are not represented here but they are present in Figure \ref{notre_architecture}.}
\label{table_reseau}
\end{table*}
\end{appendices}

\end{document}